\documentclass[%
 aapm,
 mph,%
 amsmath,amssymb,
 reprint,%
]{revtex4-1}

\usepackage{graphicx}
\usepackage{dcolumn}
\usepackage{bm}

\usepackage[mathlines]{lineno}
\modulolinenumbers[5]
\linenumbers\relax 

\begin{document}

\preprint{AAPM/123-QED}

\title[Sample title]{Limit of light coupling strength in solar cells}

\author{A. Naqavi}
\email{ali.naqavi@epfl.ch}
 \altaffiliation[Also at ]{Optics $\&$ Photonics Technology Laboratory, EPFL.}
\author{F.-J. Haug}%
\author{C. Ballif}%
\affiliation{ 
Photovoltaics and Thin-Film Electronics Laboratory
\\Ecole Polytechnique F$\acute{e}$d$\acute{e}$rale de Lausanne (EPFL)\\Rue A.-L. Breguet 2, 2000 Neuch$\hat{a}$tel,
Switzerland
}%

\author{T. Scharf}
\author{H. P. Herzig}
\affiliation{%
Optics $\&$ Photonics Technology Laboratory\\ Ecole Polytechnique F$\acute{e}$d$\acute{e}$rale de Lausanne
(EPFL)\\Rue A.-L. Breguet 2, 2000 Neuch$\hat{a}$tel, Switzerland
}%

\date{\today}

\begin{abstract}
We introduce a limit for the strength of coupling light into the modes of solar cells. This limit depends on both  a cell's thickness and its modal properties. 
For a cell with refractive index $n$ and thickness $d$, we obtain a maximal coupling rate of $2{c\sqrt{n^{2}-1}}/{d}$ where $c$ is speed of light.
Our method can be used in the design of solar cells and in calculating their efficiency limits; besides, it can be applied to a broad variety of resonant phenomena and devices. 
%

\end{abstract}

\maketitle

Enhanced light absorption  is a requirement to realize  cost-effective photovoltaics. To absorb light more efficiently, solar cells should be designed such that the incident light is coupled into them as strongly as possible \cite{krvc2003potential,haase2007thin,battaglia2011nanomoulding,moulin2012study,isabella20123,mokkapati2012nanophotonic,lanz2012photocurrent}. This can be achieved by excitation of guided or radiation modes and by benefiting from their light trapping nature \cite{ferry2008plasmonic,haug2009influence,soderstrom2010photocurrent,isabella2010modulated,naqavi2011understanding}. 

The extent to which these modes can boost absorption in a solar cell depends on how many of them the cell structure can excite and the strength of their coupling to the external radiation. Often in the literature, only the number of modes {\textemdash}or equivalently  their  density{\textemdash} is considered, which assumes that coupling is so efficient that it does not affect calculations \cite{yablonovitch1982intensity,  stuart1997thermodynamic,yu2010fundamental, haug2011resonances}. To provide more accurate values for the maximal absorption enhancement, the coupling strength of the modes should be considered as well.

In this manuscript we introduce an upper limit for the  strength of coupling  of the modes of a thin film to the outside radiation. 
For simplicity, we take into account here only guided modes  but radiation modes can be treated similarly.
For an ideally flat film, there is no coupling to external radiation but by disturbing the flat interface geometry, it is possible to couple light into the film. 
We consider a grating to provide coupling and we assume that the cell is periodic along the periodicity of this grating coupler. 
By using the coupled-mode theory explained by Yu \textit{et al.} \cite{yu2010fundamental}, we show that the maximal coupling rate depends on the thickness of the film and its modal structure. Since our aim is to find a general limit and not the exact values of the coupling coefficients, our formulation does not depend on the choice of the grating texture and even random textures may be analyzed analogously.

Time evolution of a mode in the vicinity of its resonant frequency $\omega_0$ can be described by \cite{haus1984waves}
\begin{eqnarray}
\frac{da}{dt}=\left(j\omega_{0}-N\frac{\gamma}{2}\right)a+\kappa S
\label{eq:one},
\end{eqnarray}
where $j=\sqrt{-1}$, $t$ is time, $a$ denotes the mode amplitude, $\kappa$ represents coupling to the external source $S$, $N$ is the number of output ports and the wave decay rate is $\gamma$. If we use a grating texture, each reflection order is equivalent to an output port. To ease the analysis we do not distinguish different reflection orders.  
The decay rate $\gamma$ depends on the ``internal" absorption and the  ``external" coupling; it can thus be dissociated into two parts $\gamma_{i}$ and $\gamma_{e}$. If the film is isolated from outside, the mode will be absorbed in a mean time of $\tau_{i}={2}/{\gamma_{i}}$. 
On the other hand if the film interacts with the ambient and it is lossless, the mode will couple out in a mean time of $\tau_{e}={2}/{\gamma_{e}}$. 

Intuitively one can find the minimum of $\tau_e$. 
Let us assume that an incident wave travels through the film, i.e. it couples into the film and then it couples out again. For simplicity let us also assume that the number of output ports is one so that in-coupling and out-coupling occur at the same rate. 
Since a mean time of $\tau_e$ is required for each coupling, the whole travel time will be $2\tau_e$; one $\tau_e$ for coupling in and one $\tau_e$ for coupling out. 
Therefore, by finding the minimum travel time of a wave inside the film we can obtain the minimum of $\tau_e$ and consequently the maximum of $\gamma_e$. We now apply the coupled-mode theory to test this intuition. 

If the absorption in the film is small, it can be shown that $\left|\kappa\right|=\sqrt{\gamma_{e}}$ \cite{haus1984waves}. The maximum absorption $A$ in the film occurs at $\omega=\omega_{0}$ and it can be calculated from Eq.~(\ref{eq:one}) \cite{haus1984waves, yu2010fundamental}:
\begin{eqnarray}
A&=&\left.\frac{\gamma_{i}\left|a\right|^{2}}{\left|S\right|^{2}}\right|_{\omega=\omega_{0}}
\nonumber \\&=&\left.\frac{\gamma_{i}\gamma_{e}}{\left(\omega-\omega_{0}\right)^{2}+\left(N\frac{\gamma_{e}}{2}+\frac{\gamma_{i}}{2}\right)^{2}}\right|_{\omega=\omega_{0}}
\nonumber \\&<&\frac{4\gamma_{i}}{N^{2}\gamma_{e}}
\label{eq:nine},
\end{eqnarray}
which consequently provides a bound for $\gamma_{e}$:
\begin{eqnarray}
\gamma_{e}<\frac{4\gamma_{i}}{N^{2}A}
\label{eq:ten}
\end{eqnarray}

Eq.~(\ref{eq:ten}) reveals a trade-off between the absorption and the coupling rate that eventually limits both of them. Higher absorption values represent lower coupling rates and vice versa. To ensure that we do not underestimate the coupling rate, we take the minimum possible value of absorption. We assume that the wave passes through the film only once so that the absorption over this one-pass length along the film $L$  is the minimum of absorption \cite{naqavi2013light}:
\begin{eqnarray}
A\approx\frac{\gamma_{i}}{v_{g}} L
\label{eq:eleven}.
\end{eqnarray}
where $v_{g}$ is the group velocity of the mode at resonance. From Eqs. ~(\ref{eq:ten}) and ~(\ref{eq:eleven}) one concludes that 
\begin{eqnarray}
\gamma_{e}<\frac{4v_{g}}{N^{2}L}=\frac{4}{N^{2}\tau_{t}}
\label{eq:thirteen}.
\end{eqnarray}
Here $\tau_{t}={L}/{v_{g}}$ is the wave travel time while passing through the film. 
In Eq.~(\ref{eq:thirteen}) one might use the phase velocity instead of the group velocity. 
This would give an overestimation of $\gamma_e$ but it can be useful if the group velocity is hard to define.  
For only one reflection order ($N=1$), Eq.~(\ref{eq:thirteen}) and $\gamma_{e}=2/\tau_{e}$ lead to $\tau_{e}>\tau_{t}/2$. 
This is consistent with our initial intuition that $\tau_{e,min}=\tau_{t,min}/2$.
Since $\tau_t$ depends on the modal structure of the film, the order of the mode to which the light couples and the cell thickness, one expects that $\gamma_e$ is related to these parameters similarly. Specifically, thin-film devices should support stronger coupling since they let $\tau_t$ become very small. 
Also, it should be possible to more effectively excite modes close to the light line of air because their group velocity is large.

\begin{figure}
\includegraphics{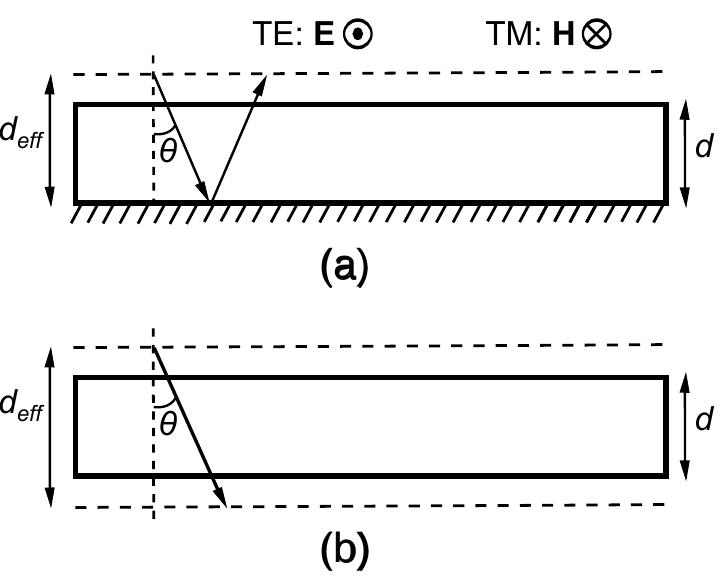}
\caption{ \label{fig_1} Schematic of an incident wave transiting a dielectric film under an angle $\theta$. The effective and the physical thickness of the guide are indicated by $d_{eff}$ and $d$. (a) The film on a back reflector, (b) The film in air. In TE/TM polarization, electric/magnetic field is normal to the incidence plane.}
\end{figure}

As a first example, consider a dielectric film of thickness $d$ and index $n$ which is sandwiched between two homogeneous half spaces of air and a perfect reflector as shown in Fig.~\ref{fig_1}(a). 
Assume that an incident wave illuminates the film from the air side and couples to a guided mode of the film associated with the internal reflection angle $\theta$. 
The incident wave passes through the film and is reflected back thus passing through the film twice. This system has one output port; based on Eq.~(\ref{eq:thirteen}) we can thus write \cite{tamir1979integrated} 
\begin{eqnarray}
\gamma_{e,max}=\frac{4v_{g}}{2d\tan\theta}=\frac{2v_{g}}{d}\frac{\sqrt{n^{2}-n_{p}^{2}}}{n_{p}}
\label{eq:fifteen}.
\end{eqnarray}
Here $n_p=\frac{c}{v_{p}}$ is the phase index where $v_{p}$ is the phase velocity and $c$ is the speed of light in air. In Eq.~(\ref{eq:fifteen}) we consider $d$ and not the effective thickness of the film, $d_{eff}$ \cite{tamir1979integrated}, since absorption occurs only inside the waveguide. 
The appearance of film thickness in the denominator of Eq.~(\ref{eq:fifteen}) reveals an advantage for thin films; they allow strong excitation provided that the mode is extremely confined, which is, e.g., the case for surface plasmons \cite{atwater2010plasmonics,munday2012light,rockstuhl2012surface}. 
The small thickness of the film will guarantee efficient coupling of light and the high confinement of the mode insures that the mode is absorbed efficiently in the film.
The group velocity and the phase index also influence $\gamma_{e,max}$ in Eq.~(\ref{eq:fifteen}) but both maximize it on the light line of air.
The maximal coupling rate of a film with one reflection order and a perfect back reflector occurs at $n_p=1$ and it is equal to ${2c\sqrt{n^{2}-1}}/{d}$.

\begin{figure}
\includegraphics{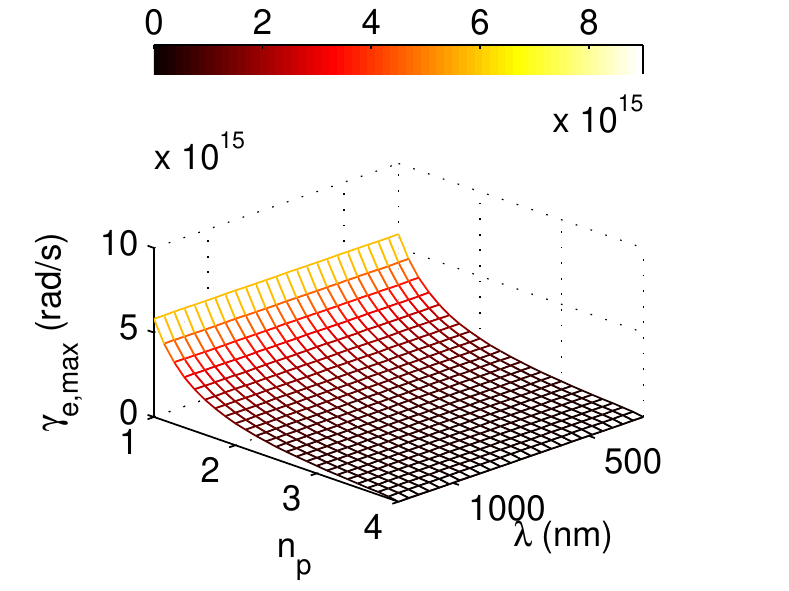}
\caption{ \label{fig_2}(Color online) $\gamma_{e,max}$ obtained by using the phase velocity approximation in Eq.~(\ref{eq:fifteen}) versus wavelength ($\lambda$) and phase index ($n_p$) for a slab with $n=4$ and $d=200\;\textrm{nm}$.}
\end{figure}

As a second example, consider a film suspended in air (as in Fig.~\ref{fig_1}(b)). 
Due to the elimination of the back reflector, this system has two ports ($N=2$) and the incident wave passes through the film only once. For simplicity, we assume that both sides of the film have the same  coupling characteristics. Eq.~(\ref{eq:thirteen}) suggests that
\begin{eqnarray}
\gamma_{e,max}=\frac{4v_{g}}{\left(2\right)^{2} d\tan\theta}=\frac{v_{g}}{d}\frac{\sqrt{n^{2}-n_{p}^{2}}}{n_{p}}\label{eq:sixteen}.
\end{eqnarray}

Figure~\ref{fig_2} shows the maximum coupling rate $\gamma_{e,max}$ for a film with refractive index $n=4$ and thickness $d=200\;\textrm{nm}$ as a function of the incident wavelength and the phase index $n_p$. 
The wavelengths correspond to the range of operation of solar cells. 
The refractive index and the thickness are chosen such that structure resembles a thin-film amorphous silicon solar cell at red wavelengths where it absorbs light weakly. 
In Fig.~\ref{fig_2}, we have applied the approximation of using the phase velocity instead of the group velocity in  Eq.~(\ref{eq:sixteen}) to obtain $\gamma_{e,max}$ regardless of polarization and band shape in the dispersion diagram. 
Therefore, Fig.~\ref{fig_2} is plotted as a continuum and the bands should be mapped on it. 
The maximum of $\gamma_{e,max}$ occurs on the light line of air ($n_p=1$) and it is equal to ${c\sqrt{n^{2}-1}}/{d}$ which is half of the value obtained in the previous example. 
Therefore, adding a backreflector is beneficial not only to increase the path length of light inside the film, but also because it allows stronger coupling. 
In this way, a solar cell's backreflector boosts the absorption of light in the cell which in turn increases the short-circuit current density and the open-circuit voltage, and consequently the efficiency \cite{miller2012strong}.

\begin{figure}
\includegraphics{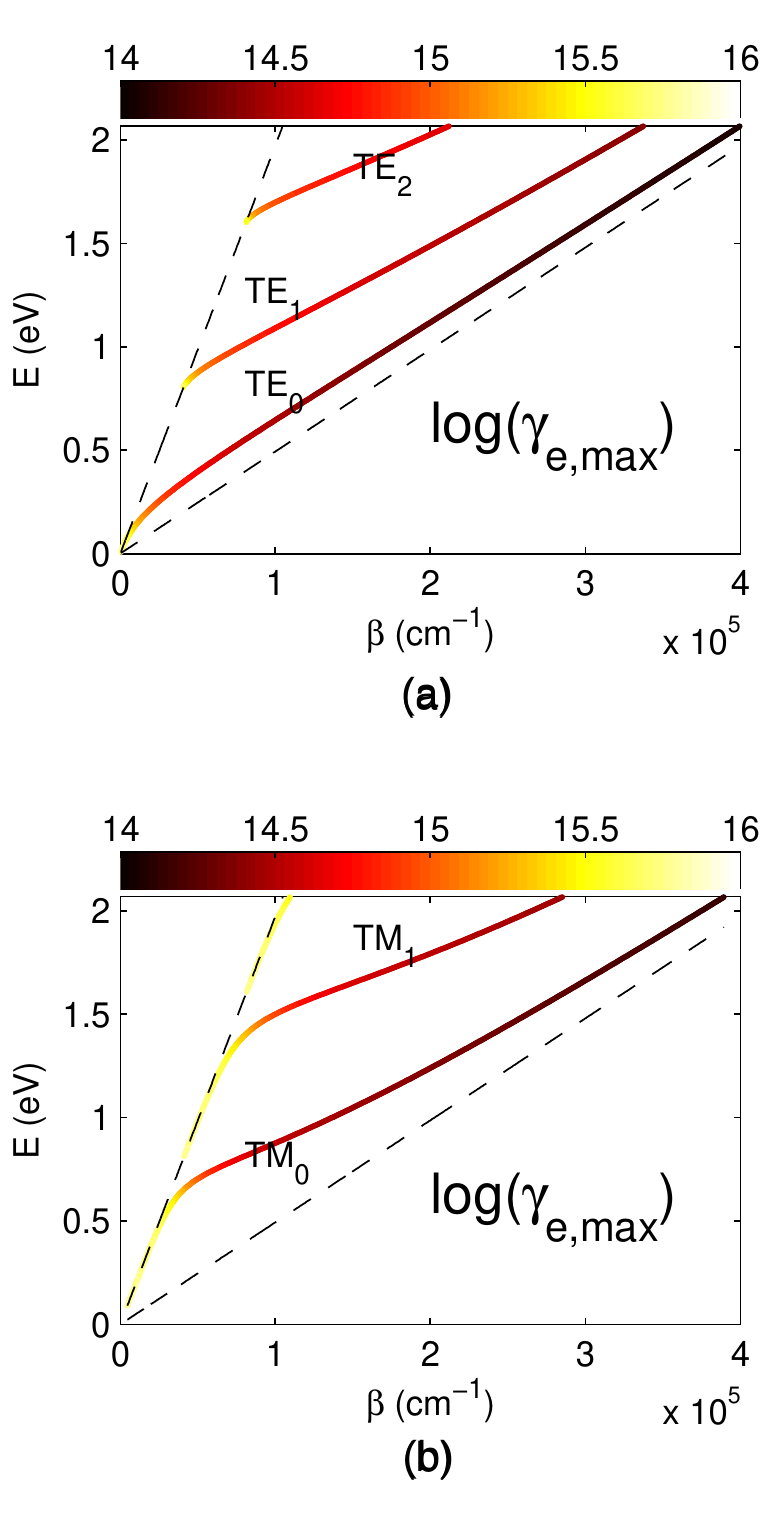}
\caption{ \label{fig_3}(Color online) $\log_{10}(\gamma_{e,max})$ for a slab with $n=4$ and $d=200\;\textrm{nm}$ for (a) TE polarization and (b) TM polarization. The dashed lines show the light lines of air and the dielectric. The unit of $\gamma_{e,max}$ is rad/s.}
\end{figure}

Figure~\ref{fig_3} (a) and (b) show $\gamma_{e,max}$ in logarithmic scale for the same film ($n=4$ and $d=200\;\textrm{nm}$) along the bands in the dispersion diagram in transverse electric (TE) and transverse magnetic (TM) polarizations. 
In TE polarization, the electric field is normal to the incidence plane (c.f. Fig.\ref{fig_1}).  
For these calculations, Eq.~(\ref{eq:sixteen}) was used with the group velocity ($v_{g}=d\omega/d\beta$).
In both polarizations, $\gamma_{e,max}$ is maximum at the onset of the modes on the light line of air. Starting from this line and moving along the bands, $\gamma_{e,max}$ decreases in agreement with the phase index approximation of Fig.~\ref{fig_2}. 
At fixed energy or propagation constant, higher-order bands provide larger $\gamma_{e,max}$ because they correspond to smaller $n_p$. 
Specifically, the TE$_2$ and the TM$_1$ modes emerge from the light line of air in the energy range where amorphous silicon is weakly absorbing, so, they have extremely high $\gamma_{e,max}$ in this range.  
Note that very close to the light line of air guided modes are weakly confined to the cell, so they cannot lead to significant absorption enhancement despite their strong excitation. 
Nevertheless, it should be possible to find a region of optimal parameters in the dispersion diagram such that both the coupling rate and the mode confinement have reasonably high values. 
One might consider this in the design of solar cells.

In summary, we derived a limit for the rate of light coupling into the modes of solar cells. 
This limit depends on the thickness of the cell, the group velocity of the mode, and where in the dispersion diagram the mode is excited. We showed a trade-off between absorption in the cell and the coupling rate. Therefore, caclulations of the maximum absorption enhancement in solar cells should be modified to consider this constraint. 
We applied our approach to solar cells but it can be used to provide limits and design guidelines for a broad variety of devices whose operation is based on resonant phenomena. 

The authors acknowledge funding from the Swiss National Science Foundation under Project No. $200021\_125177/1$.

\end{document}